\documentstyle[preprint,aps]{revtex}
\begin{document}
\draft
\def\ket#1{| #1 \rangle}
\def\bra#1{\langle #1 |}
\def\R{{\cal R}}
\def\Ha{{\cal H}_A}
\def\He{{\cal H}_{\rm e}}
\def\Hc{{\cal H}_{\rm c}}
\def\Hq{{\cal H}_{\rm q}}
\def\tto{\rightarrow}

\title{Decoherence and Efficiency of Quantum Error Correction}
\date{\today}
\author{M. Biskup$^{1,2}$, P. Cejnar$^3$, and R. Koteck\' y$^{2,1}$}
\address{$^1$Dept. of Theoretical Physics, Charles University,
V Hole\v sovi\v ck\'ach 2, 180\,00 Prague, Czech Republic\\
$^2$Center for Theoretical Study, Charles University, Jilsk\' a 1,
110\,00 Prague, Czech Republic\\
$^3$Dept. of Nuclear Physics, Charles University, V Hole\v sovi\v ck\'ach
2, 180\,00 Prague, Czech Republic}
\maketitle

\begin{abstract}
Certain physical aspects of quantum error correction are discussed for a
quantum computer ($n$-qubit
register) in contact with a decohering environment.
Under rather plausible assumptions upon the form of the computer-environment
interaction, the efficiency of a general correcting procedure is evaluated
as a function of the spontaneous-decay duration and the rank of errors
covered by the procedure.
It is proved that the probability of errors can be made arbitrarily small
by enhancing the correction method, provided
the decohering interaction is represented by a bounded operator.
\end{abstract}

\section{Introduction}

Since Shor \cite{Shor0} demonstrated that the classical factoring
problem can, in principle, be efficiently solved on a quantum computer,
a multitude of papers dealing with quantum computing have emerged.
It was soon recognized \cite{Unruh,Palma}, however, that the advantage of
implementing entangled quantum states may be spoilt by their vulnerability
to errors.
Namely, the destructive interference of the omnipresent environment leads
to an exponential loss of the  probability that the computation runs in
the desired way.
The number of runs needed to perform a successful calculation then
increases exponentially which takes one back to the purview of complexity
problems.

These pessimistic views have abated to a certain extent since the first work
pioneering quantum error correction appeared \cite{Shor1}.
It immediately  became a subject of fascination that something as
delicate to handle as quantum state can be mended without knowing any
particulars about it.
In view of that, a variety of error correcting codes \cite{ZOO1}
and related complexity requirements \cite{Laflamme,Gottesman,Steane}
have been thoroughly discussed.

The basic idea of quantum correcting is fairly simple.
The computational state is encoded, by introducing redundancy, into
a more robust one, which can then be rectified, provided only errors
from a certain {\it sub-}class have occurred.
It is subsequently argued that the remaining errors appear with a small
probability, which is a key prerequisite for the proof of correction
effectiveness.
A typical example of the above sub-class are single-qubit errors.
In this case, simultaneous errors on more than one qubit are expected to
conform to the law of independent probabilities, and, therefore, to be
dominated by the single-qubit errors.
If the latter errors can be put away, the system is indeed much less
susceptible to perturbations.

While quantum codes and related topics have been well explored during
over a year of their existence, little has been said on the genuine
physical aspects of the quantum error correction, though some of them
certainly deserve our attention.
Namely, despite all correction methods resemble the watch-dog
stabilization (discussed, e.g., in \cite{Zurek2}), there is an
important difference:
For the watch-dog stabilization (in the ideal case) to function, corrections
have to be repeated at an infinite rate.
But, in reality, there is always a principal bound on the correction
frequency (apart from others, for the quantum computation must not be
interfered with)---the exponential decay of the original state cannot be
avoided.
Of course, this does not necessarily imply that the error correction
brings no profit.
In order to estimate the correction efficiency, however, the temporal
aspects of the correction process have to be carefully scrutinized.
As shown in the following example, this cannot be done irregardless
the particulars of the decohering interaction.

Consider a simple correction scheme capable of eliminating single- and
double-qubit amplitude errors (i.e., $\ket{0}\leftrightarrow\ket{1}$
flips---see sect.\,II):
States of a single qubit are encoded into states of a qubit pentet,
$\ket{0}\to\ket{00000}$ and $\ket{1}\to\ket{11111}$, and the correction
is represented by the transformation $\ket{b_1b_2b_3b_4b_5}\to\ket{bbbbb}$
on the encoded states, where $b$ is equal to $b_i$ that occurs at least
three times in the initial state (majorization rule).
Let us try to compare the stabilization effect of the above correction
under two different error-producing interactions: $H^{(1)}=\sum_l\hbar
\omega_l\sigma_x^l$ and $H^{(2)}=\sum_{k\neq l}\hbar\omega_{kl}
\sigma_x^k\sigma_x^l$, where $\sigma_x^i$ is the first Pauli matrix
($\sigma_x:\ket{0}\leftrightarrow\ket{1}$) operating on the $i$-th qubit.
As both the interactions cause only amplitude errors, the correction
procedure is, in principle, applicable.
The corresponding evolution operators have the form:
$U^{(1)}(t)=\prod_{l}\bigl[ \cos(\omega_lt)+i\sigma_x^l\sin(\omega_lt)
\bigr]$ and $U^{(2)}(t)=\prod_{k\ne l}\bigl[ \cos(\omega_{kl}t)+
i\sigma_x^k\sigma_x^l\sin (\omega_{kl}t)\bigr]$.
By breaking the products into parts, it is evident that the error
proliferation is much faster for the second hamiltonian.
Namely, for short times $t$, the errors which cannot be cured by our
procedure (i.e., the terms in $U(t)$ corresponding to three- and
higher-qubit errors) are of order $t^3$ in the first case and $t^2$ in
the second one.
To achieve the same level of stabilization for both interactions, the
corrections in the second case have to be repeated at much higher rate.

The plan of the paper is as follows.
In the second section we briefly recall the basic facts on decoherence
and quantum error correction.
The third section is devoted to the evaluation of the correction efficiency,
under a certain mild assumption on the form of decohering interaction.
In the last section we present conditions under which a rigorous meaning
can be given to our formulas, and discuss their applicability to
realistic situations.
In particular, we prove that the quantum computer can be stabilized
if the decohering interaction is represented by a bounded operator.

\section{Decoherence and Correction by Codes}

Decoherence in quantum systems is, according to the current operationalistic
point of view\cite{ZurekPT}, induced by coupling the system to its
environment.
As a result, the states of the computer become entangled with the
environmental ones, the fact that has a deadly impact on quantum
superpositions.
Formally, decoherence is described by the chain
\begin{equation}
\varrho_0\longrightarrow D_0=\varrho_0^{\rm e}\otimes\varrho_0
\stackrel {t}{\longrightarrow} D_t=U(t)D_0 U^{\dag}(t)\longrightarrow
\varrho_t={\rm Tr}_{\He}D_t,
\label{decoheremce}
\end{equation}
where $\varrho_0$ and $\varrho_t$ are the computer states (density
matrices) at time 0 and $t$, respectively, and $\varrho_0^{\rm e}$ is
the initial environmental state.
Here the leftmost arrow represents enlarging of the computer's Hilbert
space $\Hc$ by the environmental degrees of freedom, the middle one stands
for the joint computer-environment evolution (resulting typically
in a non-product density matrix $D_t$), and the rightmost arrow reflects
our ignorance of the environment, expressed in terms of the partial trace
over the environmental Hilbert space $\He$.
Despite the unitarity of the joint evolution $D_0\rightarrow D_t$, the
full transformation $\varrho_0\rightarrow\varrho_t$ need no longer be
unitary.

During the run of a calculation, the computational state faces a sequence
of neatly timed unitary transformations.
In the meantime, when quantum gates are being readied for the next
computational step, the quantum registers containing qubits are falling
victim to the harsh intervention of the environment.
Consequently, to study decoherence effects in the quantum computer, one
should primarily be interested in qubits suspended in the registers that
are exposed just to the environmental interaction.

The simplest quantum register contains only one qubit (we denote the
single-qubit Hilbert space by $\Hq$).
The action of the overall evolution operator $U(t)$ can be formalized by
the following equations
\begin{equation}
U(t)|e_i\rangle|0\rangle=|g_i\rangle|0\rangle+|l_i\rangle|1\rangle,
\quad U(t)|e_i\rangle|1\rangle=|u_i\rangle|0\rangle+|m_i\rangle|1\rangle.
\label{inter}
\end{equation}
Here $|e_i\rangle$ is an orthonormal basis of $\He$ (for simplicity
we suppose $\varrho_0^{\rm e}=\sum_i w_i|e_i\rangle
\langle e_i|$), and $|g_i\rangle$, $|l_i\rangle$, $|u_i\rangle$ and
$|m_i\rangle$ are some unknown environmental states (neither normalization
nor orthogonality is required) containing $t$ as an implicit variable.
The crucial observation \cite{Shor1}, leading ultimately to the correcting
codes, is that the time-dependence of the evolution operator $U(t)$
can be totally embodied in the environmental Hilbert space $\He$.
Namely, $U(t)$ admits a trivial factorization
$U(t)=\sum_{\mu} U_{\mu}(t)\otimes Q_{\mu}$, where $U_{\mu}(t)$ and
$Q_{\mu}$ ($\mu=0,1,2,3$) act on the environmental and qubit states,
respectively.
Let $|\psi\rangle=\alpha|0\rangle+\beta|1\rangle$.
The explicit form of $U_{\mu}(t)$ and $Q_{\mu}$ can be deduced
from the formula
\begin{eqnarray}
U(t)|e_i\rangle|\psi\rangle & = & \frac{|g_i\rangle+|m_i\rangle}{2}
(\alpha|0\rangle+\beta|1\rangle)+\frac{|g_i\rangle-|m_i\rangle}{2}
(\alpha|0\rangle-\beta|1\rangle)+\nonumber
\\
&&\phantom{mmmmmmmmm}+\frac{|l_i\rangle+|u_i\rangle}{2}
(\beta|0\rangle+\alpha|1\rangle)+\frac{|l_i\rangle-|u_i\rangle}{2}
(-\beta|0\rangle+\alpha|1\rangle) \nonumber=\\
& = & |a_i\rangle\bbox{1}|\psi\rangle+|b_i\rangle\sigma_z|\psi\rangle+
|c_i\rangle\sigma_x|\psi\rangle+|d_i\rangle(-i\sigma_y)|\psi\rangle.
\label{separ}
\end{eqnarray}
Here $\sigma_x$, $\sigma_y$, and $\sigma_z$ are Pauli matrices in the basis
$|0\rangle$ and $|1\rangle$.
The expression (\ref{separ}) has led to a \lq\lq discrete\rq\rq\
classification of quantum errors \cite{Laflamme} into amplitude errors
($\sigma_x$), phase errors ($\sigma_z$), and combined amplitude-phase errors
$(-i\sigma_y=\sigma_z\sigma_x)$; $\bbox{1}$ represents no error.
In view of (\ref{separ}), if the three classes of errors can be corrected,
then any (even unknown) single qubit state can resist perturbations.

An error decomposition analogous to (\ref{separ}) is valid also
for $n$-qubit systems. If $U(t)$ is an evolution operator (even non-unitary)
on $\cal H=\He\otimes\Hc$, with $\Hc=\Hq^{\otimes n}$, then
\begin{equation}
U(t)=\sum_{\{ \mu_i\} }U_{\{ \mu_i\} }(t)\sigma^1_{\mu_1}
\sigma^2_{\mu_2}\dots \sigma^n_{\mu_n},
\label{gensepar}
\end{equation}
where the sum runs over all $n$-collections $\{ \mu_i\}_{i=1}^n$
of indices $\mu_i\in\{ 0,1,2,3\} $ and $U_{\{ \mu_i\}}$ are operators
on $\He$, corresponding to the respective dynamics in the qubit sector.
Here we have made use of the shorthand notation
\begin{equation}
\sigma^k_{\mu} = \underbrace{\bbox{1}\otimes
\dots\otimes\bbox{1}}_{k-1}\otimes
\,\sigma_{\mu}\otimes
\underbrace{\bbox{1}\otimes \dots\otimes \bbox{1}}_{n-k},
\label{sigmamu}
\end{equation}
where $\sigma_{\mu}$=$(\bbox{1},\vec\sigma)$.
The decomposition (\ref{gensepar}) is unique since it can be inverted
to yield $U_{\{ \mu_i\} }(t)=2^{-n}{\rm Tr}_{\Hc}\bigl(U(t)
\sigma^1_{\mu_1}\sigma^2_{\mu_2}\dots\sigma^n_{\mu_n}\bigl)$.
Now, it is meaningful to say that an error of {\it rank} $k$ has
occurred, if only collections with exactly $k$ non-zero indices
$\mu_i$ contribute to (\ref{gensepar}).

Error correction by codes relies upon the possibility to encode single
logical states $\ket{0}$ and $\ket{1}$ into  specific superpositions
$\ket{\bar 0}$ and $\ket{\bar 1}$ of compound $n$-qubit logical states
(so-called code-words from $\Hq^{\otimes n}$).
If the states $\ket{\bar 0}$ and $\ket{\bar 1}$ are distant enough from
each other, where the Hamming distance turns out to be an appropriate
measure for these purposes, the histories of a certain sub-class of
errors can be traced back and subsequently remedied.
Let $k$ be fixed.
Then it turns out that for some $n$ satisfying the quantum Hamming and
Gilbert-Varshamov bounds \cite{EM},
\begin{equation}
\sum_{l=0}^{k}{n\choose l}3^l
\le 2^{n-1}\le
\sum_{l=0}^{2k}{n\choose l}3^l,
\label{Hambound}
\end{equation}
there exists a code into $n$ qubits capable of rectifying all errors
up to rank $k$.
The above inequalities follow from the requirement that the Hilbert space
of code-words is large enough to allow both, the original information
contained in $\varrho_0$ (cf. formula \ref{decoheremce}) as well as the
way it has been altered by the above sub-class of errors, to be deciphered
from the register density matrix $\varrho_t$.
It will be important for us to observe the asymptotic behaviour of the
above bounds as $n\tto\infty$.
Taking at the same time $k/n\to x$, the formula (\ref{Hambound}) boils
down to
\begin{equation}
x\ln 3-\ln\bigl[ x^x (1-x)^{1-x}\bigr]
\le\ln 2\le
2x\ln 3 - \ln\bigl[(2x)^{2x}(1-2x)^{1-2x}\bigr].
\label{asympbound}
\end{equation}
Since both the inequalities are satisfied for a nonvanishing argument
($x\in[x_0,2x_0]$ with $x_0\approx 0.1$), we see that the number of errors
$k$ that can be controlled grows roughly linearly with the size $n$ of the
code.

The repair of defects is standardly thought to be done by means of
auxiliary qubits, so-called ancillas.
Before the correction procedure is initiated, the ancillas are prepared
in a fixed state $|A\rangle$ (this is important as it implies zero
entropy input---disorder can be transferred to the ancilla Hilbert space
$\Ha$).
When they are brought into contact with the computational qubits, the
corresponding type of error is re-printed in the ancilla state and
subsequently rectified in the computational sector.
Formally, to represent this operation, one introduces \cite{Knill} a
recovery operator $\R$, acting on the product space  $\Hc\otimes \Ha$.
As can be easily shown, $\R$ can be made unitary \cite{Knill,Bennett}, which
allows for the coherent inclusion of the error correction into the
computational algorithm.

\section{Correction efficiency}

It is intuitively clear that the correction procedures function only
when the stored computational state has not departed very far from the
initial one.
In this section we would like to discuss this common supposition
quantitatively.
In order to do that, let us recall \cite{Schumacher} the definition of the
fidelity
functional ascribed to the channel decoherence\,+\,correction
(represented by the operator $\R U$).
Namely, the fidelity functional $F_{\psi}(t)$ is the probability that the
initial state $|\psi\rangle$ passes through the channel intact.
If the environment is originally in the state $\varrho_0^{\rm e}$, then
\begin{equation}
F_{\psi}(t)={\rm Tr}\Bigl[\R U(t)(P_{\psi}\otimes\varrho_0^{\rm e}\otimes
P_A)U^{\dag}(t)\R^{\dag}P_{\psi}\Bigr].
\label{fidelity}
\end{equation}
Here $P_{\psi}=|\psi\rangle\langle\psi|$ is the projector
onto the state $|\psi\rangle$ and, similarly,  $P_A=|A\rangle\langle A|$.
Without being explicitly marked out, the ancilla is not
to be affected by the evolution under $U(t)$.
Consequently, error $E(t)$ of the respective code is defined
\cite{Knill} by $E(t)=\sup_{\ket{\psi}}E_{\psi}(t)$, where
$E_{\psi}(t)=1-F_{\psi}(t)$ is the error functional.
It is worth noting that as we suppose $\R$ to be
unitary, $E_{\psi}$ is expressed by the same formula as $F_{\psi}$ with
only the last projector $P_{\psi}$ replaced by $\bbox{1}-P_{\psi}$.

Assume now $n$ qubit registers being in the state $|\Psi\rangle$.
The registers are exposed to the environmental intrusion, described
by an interaction hamiltonian $V$.
In the following, we shall be concerned with the class of non-contact
interactions, i.e., those for which $V$ takes the form
\begin{equation}
 V=\sum_{l=1}^n\sum_{\mu=1}^3  h_{\mu}^l\otimes \sigma_{\mu}^l,
\label{ham}
\end{equation}
where $h_{\mu}^l$ are some self-adjoint operators on the environmental
Hilbert space (the term with $\mu=0$ is absent in equation (\ref{ham})
as it can be incorporated into the free hamiltonian---see below).
This assumption is justifiable if all inter-qubit communications
are mediated by some (quasi-) particle external fields.
For instance, in the linear ion-trap computer single qubits interact
by exchanging photons and phonons, so the interaction is indeed a
non-contact one.

In the interaction picture, the evolution operator $U(t)$ satisfies
the Schr\"odinger equation
\begin{equation}
i \frac {d}{dt} U(t) = V(t) U(t),
\label{intpict}
\end{equation}
where $V(t)$ is the freely evolved interaction hamiltonian,
$V(t)=e^{iH_0t}Ve^{-iH_0t}$ (we set $\hbar=1$).
As the free hamiltonian is not supposed to induce interaction between
qubits ($H_0$ is the sum of environmental and single-qubit terms),
the free evolution of $\sigma$'s in (\ref{ham}) can be incorporated into
the evolution of $h$'s.
Namely, taking into account that $\sigma_{\mu}^l(t)=\sum_{\nu}
f^l_{\nu\mu}(t)\sigma_{\nu}^l$ (where $f^l_{\nu\mu}(t)$ are some functions
of time), we can put the operator $V(t)$ to the same form as $V$, only with
$h_{\mu}^l$ being replaced by $\sum_{\lambda}f_{\mu\lambda}^l(t)
h_{\lambda}^l(t)$.

Let the environment be in the state $|e_i\rangle$.
Then the joint time evolution of $|e_i\rangle|\Psi\rangle$ in the
interaction picture can be written
\begin{eqnarray}
U(t)|e_i\rangle|\Psi\rangle &=& |e_i\rangle|\Psi\rangle -i
\int_0^t ds\,  V(s) U(s) |e_i\rangle|\Psi\rangle  \nonumber=\\
&&\phantom{xxxxxxxx}=
\sum_{l=0}^k \frac {1}{l!} T\biggl(-i\int_0^t ds\, V(s) \biggr)^l
 |e_i\rangle|\Psi\rangle +\nonumber\\
&+& (-i)^{k+1}\int_0^t ds_1 \int_0^{s_1} ds_2\dots
\int_0^{s_k} ds_{k+1}V(s_1)V(s_2)\dots
V(s_{k+1})U(s_{k+1}) |e_i\rangle|\Psi\rangle ,
\label{casev}
\end{eqnarray}
where $T$ is the time-ordering operator and $k$ is chosen to be precisely
the rank of errors we intend to correct.
The second equality has been obtained by iterating the preceding one
$k$-times, while tacitly supposing that $U(t)$ is sufficiently
differentiable when applied to $|e_i\rangle|\Psi\rangle$.
Now let the ancilla qubits (initially in the state $|A\rangle$) join the
evolution and carry out the correction $\R$.
Since the correction procedure cures all errors up to rank $k$, the
first term in (\ref{casev}) is thoroughly reverted to the state
$|\Psi\rangle$ in the computational sector and, therefore,
brings no contribution to the error functional $E_{\Psi}(t)$
(due to the presence of $\bbox{1}-P_{\Psi}$). This enables us to write
\begin{eqnarray}
E_{\Psi}(t)&=& \int_0^t ds_1\dots\int_0^{s_k}ds_{k+1} \int_0^t ds'_1 \dots
\int_0^{s'_k}ds'_{k+1} \nonumber\\
&&\phantom{x}{\rm Tr}\bigl[ \R V(s_1)\dots V(s_{k+1}) U(s_{k+1})
D_{\Psi,A} U^{\dag}(s'_{k+1})V(s'_{k+1})\dots V(s'_1)
\R^{\dag} (\bbox{1}-P_{\Psi})\bigr ],
\label{error}
\end{eqnarray}
with $D_{\Psi,A}=P_{\Psi}\otimes\varrho_0^{\rm e}\otimes P_A$.
We have the first result:
{\it If the correcting method is capable of remedying all errors up to
$k$, then $E_{\Psi}(0)=E'_{\Psi}(0)=\dots= E^{(2k+1)}_{\Psi}(0)=0$
for all initial states $|\Psi\rangle$}.
Moreover, the behaviour for short times is given by the relation
\begin{equation}
E_{\Psi}(t)=\frac{t^{2k+2}}{(k+1)!^2}
\sum_{\{ l_i\} }\sum_{\{ l'_i\} } {\rm Tr}\bigl [
\R  V^{l_1}\dots V^{l_{k+1}}D_{\Psi,A}
V^{l'_{k+1}}\dots V^{l'_1}\R^{\dagger}(\bbox{1}\!-\!P_{\Psi})
\bigr ] + {\cal O}(t^{2k+3}),
\label{aprxerror}
\end{equation}
where the sums run over all ordered $k+1$-subsets of $\{ 1,2,\dots,n\} $,
and $V^{l}=\sum_{\mu=1}^3 h^l_{\mu}(0)\otimes\sigma^l_{\mu}$.
The formula (\ref{aprxerror}) is obtained by substituting $V(s)
\approx V(0)$ and $U(s)\approx\bbox{1}$ for small times $s$ in
(\ref{error}).
As the errors incurred in (\ref{aprxerror}) by the chains of $V$'s are
of rank $k+1$, they are not all likely to be corrected for a generic
environmental interaction, unless the correction method encompasses also
higher order errors.
Thus, $E^{(2k+2)}_{\Psi}(0)\not=0$ and (\ref{aprxerror}) describes the
true behaviour of $E_{\Psi}$ for short times.

We see that, indeed, the more enhanced codes are applied the slower the
errors escalate.
It is worth noting that the formula (\ref{aprxerror}) generalizes
the standard treatment of the quantum watch-dog effect.
In the latter, no error correction is employed, therefore $k=0$, which
is in accord with the obtained time dependence ($E_{\psi}(t)\sim t^2$).
However, as opposed to the watch-dog effect, quantum error correction by
codes requires no knowledge of the initial state.

\section{Stabilization by Correction}

We have seen in the previous section that the ancilla based correction
brings about polynomial slow-down of error propagation, provided the mild
condition (\ref{ham}) upon the form of the interaction is satisfied.
In practice, however, this does not imply that decoherence can be stopped
from penetrating into the computer.
Namely, as already mentioned, since the frequency at which the
rectification is repeated cannot be made arbitrarily large, even
the encoded quantum information in the register subjected to periodic
corrections decays {\it exponentially} with time, with only the decay
rate reduced.
In view of these remarks, the following concept of stabilization is more
appropriate:
A correction procedure, depending on a discrete parameter $n$, qualifies to
be stabilizing, if there is a range of times for which the probability of
errors can be made arbitrarily small by varying the parameter $n$.
The latter parameter represents the comprehensiveness of the correction
method and is exemplified by the length of code-words in our case.

We would like to clarify when a computer register can be stabilized in the
above sense.
In particular, we prove the following statement: {\it If the decohering
interaction has the form (\ref{ham}), with the operators $h_{\mu}^l$
uniformly bounded, then there exists an error procedure stabilizing
quantum computer against error proliferation}.
It should be noted that we make no particular assumptions on the free
hamiltonian, apart from the natural self-adjointness requirement ensuring
the existence of $e^{-iH_0t}$.
For the proof, notice that in this setting all the above formulas
can be given a good mathematical meaning.
Consequently, we can derive the inequality
\begin{equation}
E(t)\le \frac{t^{2k+2}}{(k+1)!^2} \Vert V \Vert^{2k+2}
\label{bound}
\end{equation}
by directly utilizing the boundedness assumption in (\ref{error})
($\Vert\cdot\Vert$ is the standard operator norm).
If $\Vert h_{\mu}^l \Vert\le C$ for all $\mu$ and $l$, then, in the
regime $k,n\to\infty$, we get
\begin{equation}
E(t)\le {\cal O}(1)
\biggl[t C e\frac{n}{k}\biggr] ^{2k+2} \ ^{<}_{\sim}\
\biggl[ t\frac{Ce}{x_0}\biggr]^{2x_0 n},
\label{bound2}
\end{equation}
where $x_0$ stands for the lower bound on the asymptotic value of $k/n$
(see (\ref{asympbound})).
Hence, if $t<x_0/Ce$, the error $E(t)$ tends to zero exponentially
fast as $n$ increases.
Consequently, by choosing a sufficiently enhanced encoding, the fidelity
of the decoherence+correction channel outstrips every bound $1-\delta$
(with $\delta>0$) and the quantum information is, indeed, well stabilized.

Unfortunately, it turns out that the genuine environmental interaction
cannot be represented by a bounded operator.
Nevertheless, the high-energy environmental states are usually strongly
suppressed due to low temperature $T$.
Since the characteristic scale of the interaction energy $h_{\mu}^l$
is of order $k_{\rm B}T$, one is led to the rough estimate $t<t_T\sim 1/
k_{\rm B}T$ for the regime of applicability of the quantum error
correction.
Hence, the threshold of the thermal regime $t_T$ (cf.\,\cite{Unruh})
probably sets an absolute bound beyond which no correction is of
substantial help.

The rigorous treatment of the unbounded case in this generality falls
beyond the current level of mathematical quantum theory.
It is not so difficult to control the formula (\ref{error}) if
certain plausible assumptions on the interaction $H_0+V$ can
be made.
In particular, one goes from the description in terms of the
Dyson series to the description by the Born series \cite{Tip}.
It is not clear, however, whether it is possible to derive an analogue
of the bound (\ref{bound}), in particular, with the factor $(k+1)!^2$
in the denominator (notice that this factor is absolutely necessary to
compensate the natural extensive behaviour of the operator $V$).
We plan to address these topics in a future work.

The restriction to non-contact interactions is,
actually, not essential. Our treatment can be applied also to contact
hamiltonians
(with multiple-$\sigma$ terms), leading, however, to correspondingly
weaker results concerning the correction efficiency.

\end{document}